\documentclass{webofc}
\usepackage[varg]{txfonts}   
\begin{document}
\title{Extending to the Submillimeter Universe with the CCAT Observatory}

\author{\lastname{E.~M.~Vavagiakis}\inst{1,2}\fnsep\thanks{eve.vavagiakis@duke.edu}
        \lastname{for the CCAT Collaboration}
}

\institute{Department of Physics, Duke University, Durham, NC 27710, USA
\and
           Department of Physics, Cornell University, Ithaca, NY 14853, USA 
}

\abstract{
  The CCAT Observatory's Fred Young Submillimeter Telescope, a novel, high-throughput, 6-meter aperture telescope, is scheduled for first light in 2026. Located at 5600 m on Cerro Chajnantor in the Chilean Atacama Desert, the CCAT site enables unprecedented submillimeter measurement capabilities, fully overlapping with millimeter-wave surveys like the Simons Observatory. CCAT will address a suite of science goals, from Big Bang cosmology, star formation, and line-intensity mapping of cosmic reionization, to galactic magnetic fields, transients, and galaxy evolution over cosmic time. We highlight CCAT's science goals with Prime-Cam, a first generation science instrument for the Fred Young Submillimeter Telescope. Prime-Cam will field over 100,000 kinetic inductance detectors across seven instrument modules to enable over ten times faster mapping speed than previous submillimeter observatories in windows between 1.4 – 0.3 mm (220 – 850 GHz). We give an instrument summary, discuss the project status, and outline preliminary plans for early science.
}
\maketitle
\section{Introduction} \label{intro}

The CCAT Collaboration, an international partnership of institutions led by Cornell University, is building an observatory at a unique site in the Chilean Atacama Desert to pursue some of the most fundamental questions in astrophysics and cosmology.\footnote{www.ccatobservatory.org} In the coming years, experiments such as the Simons Observatory (SO) \cite{so} will be offering us unprecedented sensitivity at millimeter wavelengths, opening new windows onto the early and evolving universe through measurements of the cosmic microwave background (CMB). By building the Fred Young Submillimeter Telescope (FYST, pronounced ``feast"), CCAT will conduct a wide-field submillimeter survey from the Atacama, the first of its kind. This highly complementary survey will sensitively measure foregrounds that would otherwise limit millimeter surveys in our systematics-dominated era of cosmology. FYST's capabilities will also provide new insights into the Epoch of Reionization, unveil galaxy evolution over cosmic time, measure transients, and study our Milky Way in velocity-resolved submillimeter spectroscopy \cite{ccatscience}. 

\section{The Fred Young Submillimeter Telescope} \label{fyst}

FYST is a 6-meter aperture telescope designed to conduct wide-field observations across the millimeter to submillimeter range (100 GHz – 1.5 THz) \cite{parshley2018}. The FYST design, a modified crossed-Dragone telescope with aluminum-tiled primary and secondary mirrors supported by a carbon fiber structure, was also adapted for the SO Large Aperture Telescope (LAT) \cite{niemackfyst,solat}. FYST adopted tighter tolerances than the LAT for submillimeter science, and will deliver a high surface accuracy (< 10.7 $\mu$m HWFE) with low emissivity (< 2$\%$). FYST provides a coma-corrected, diffraction-limited field of view of 8$^{\circ}$ in diameter at 2 mm (100 GHz) to 2$^{\circ}$ in diameter at 0.35 mm (850 GHz). This wide field of view allows the telescope to accommodate over 100,000 diffraction-limited beams at wavelengths below 1 mm, supporting instruments with unprecedented submillimeter sensitivities.

Located at an altitude of 5600 m on Cerro Chajnantor in the Atacama Desert (near the SO site, but at a higher elevation), FYST benefits from some of the best atmospheric conditions on Earth for submillimeter astronomy \cite{astro2020}. Designed and built by CPI Vertex Antennentechnik, GmbH, Germany, FYST is currently under construction at the Cerro Chajnantor site. The mirrors have shipped, and installation is scheduled for first light in 2026. FYST and its instruments are optimized for ongoing remote operation under harsh conditions. The system is built to support a nominal 15-year lifetime and includes three instrument bays, providing a robust platform for current and future instruments to address a broad range of astrophysical and cosmological science goals. FYST will field two first-generation science instruments: Prime-Cam (Sec. \ref{primecam}) and the CCAT Heterodyne Array Instrument (CHAI) \cite{chai}. Prime-Cam will sit in the primary instrument space and be allocated 75$\%$ of the available observing time over the first five years (\textgreater 10,000 hours). CHAI, a 2$\times$64 pixel, dual frequency band, high-resolution spectrometer designed to study the cycle of interstellar matter in galaxies, is being constructed by the University of Cologne and will use the remaining 25$\%$ of the observing time \cite{chai}.

\begin{figure}[h]
\centering
\includegraphics[scale=0.245]{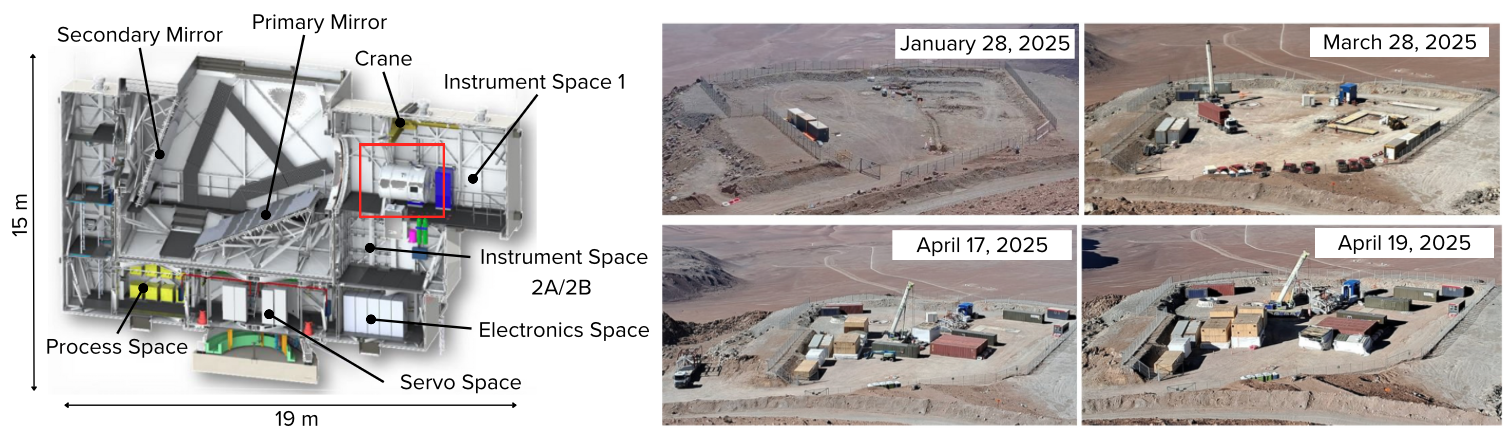}
\caption{Left: A cross-section rendering of the Fred Young Submillimeter Telescope, showing 6 m primary and secondary mirrors, and instrument, electronics, and process spaces. Prime-Cam, highlighted in red, will sit in Instrument Space 1, and be assembled using the space's crane. Right: The 5600 m elevation FYST site on Cerror Chajnantor has seen significant development, and FYST is currently under construction for first light in 2026. 
}
\label{fig:fyst}     
\end{figure}

\vspace{-0.5cm}
\section{Science Goals with Prime-Cam on FYST} \label{science}

\begin{figure}[h]
\centering
\includegraphics[scale=0.168]{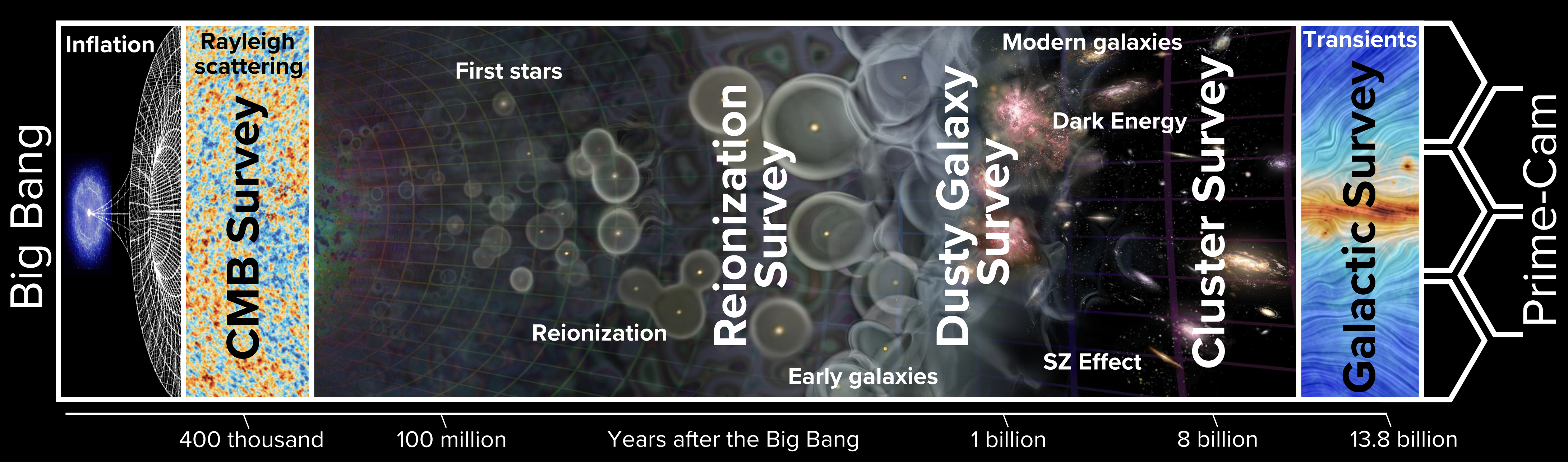}
\caption{A timeline of the universe's evolution, highlighting Prime-Cam surveys and science goals. The CMB Survey will target CMB foregrounds, aid in the search for primordial gravitational waves to study inflation, and search for Rayleigh scattering of the CMB. The Reionization Survey will map Epoch of Reionization with C+ and CO line intensity mapping. The Dusty Galaxy Survey will trace galaxy formation through Cosmic Noon. The Galaxy Cluster Survey will study galaxy and cluster formation and the Sunayev-Zeldovich effects. The Galactic Survey will characterize our Milky Way's magnetic fields and pursue galactic polarization science. Finally, a transient search campaign will pursue submillimeter transient measurements over a wide range of time scales.
}
\label{fig:science}     
\end{figure}

Prime-Cam will field up to seven independently optimized instrument modules to address a broad suite of science goals, bridging astrophysics and cosmology and spanning cosmic history from the Big Bang to the present day (Fig. \ref{fig:science}). For a detailed overview, see \cite{ccatscience}. A brief overview of the main areas of focus is outlined here. 

Probing the earliest moments of our universe's history, a central goal in cosmology is to test models of inflation by searching for signals of primordial gravitational waves in the polarization of the cosmic microwave background (CMB). Foreground emission is also polarized, making the measurement of these signals challenging. Prime-Cam on FYST will be uniquely suited to characterizing and removing this foreground emission from this and other signals in the CMB. Prime-Cam will also leverage broad frequency coverage to pursue a first measurement of Rayleigh scattering of the CMB. The CMB is also a backlight for studies of large scale structure and galaxy clusters, and Prime-Cam's frequency coverage will effectively map the Sunayev-Zeldovich effect with a better handle on systematics and component separation than with CMB data alone. 

Prime-Cam will field two spectroscopically-enabled instrument modules to trace the history of star and galaxy formation and the growth of large scale structure during the Epoch of Reionization (6 $\lesssim$ z $\lesssim$ 20) by mapping redshifted [CII] and [OIII] fine-structure line emission from early galaxies. At redshifts at and beyond the peak of cosmic star-forming history (``Cosmic Noon", 1 $\lesssim$ z $\lesssim$ 3), Prime-Cam's frequency coverage will be able to map the emission from dusty star-forming galaxies to reveal the universe's obscured structure growth to trace galaxy assembly and star formation from the Epoch of Reionization until when dark energy begins to have important impact on expansion. FYST will also search for submillimeter transients, such as supernovae, gamma-ray bursts, merging neutron stars, and tidal disruption events. Our understanding of these energetic phenomena will benefit from Prime-Cam's spectral coverage. More locally, Prime-Cam will trace magnetic fields in our own Milky Way from kpc to the sub-pc scales at which individual stars form.

\vspace{-0.5cm}
\section{Prime-Cam Overview and Status} \label{primecam}

Prime-Cam is a 1.8 m diameter receiver designed to fill the central 4.9$^{\circ}$ of FYST's field of view (FoV) with seven independently optimized instrument modules, each with a 1.3$^{\circ}$ FoV \cite{primecam2024,primecam2020,primecam2018,ccatscience}. Two of Prime-Cam's instrument modules will be spectroscopically enabled for line intensity mapping, and five will be broadband, polarization-sensitive instruments. Together, the seven modules will deploy over 100,000 kinetic inductance detectors (KIDs, Sec. \ref{sec:detectors}) covering 220 -- 850 GHz, the largest deployment of KIDs to date. The Prime-Cam instrument module designs were designed based on the SO Large Aperture Telescope Receiver (LATR) optics tube designs \cite{zhuSO}, and are discussed in Sec. \ref{sec:instmod}.

\begin{figure}[h]
\centering
\includegraphics[scale=0.19]{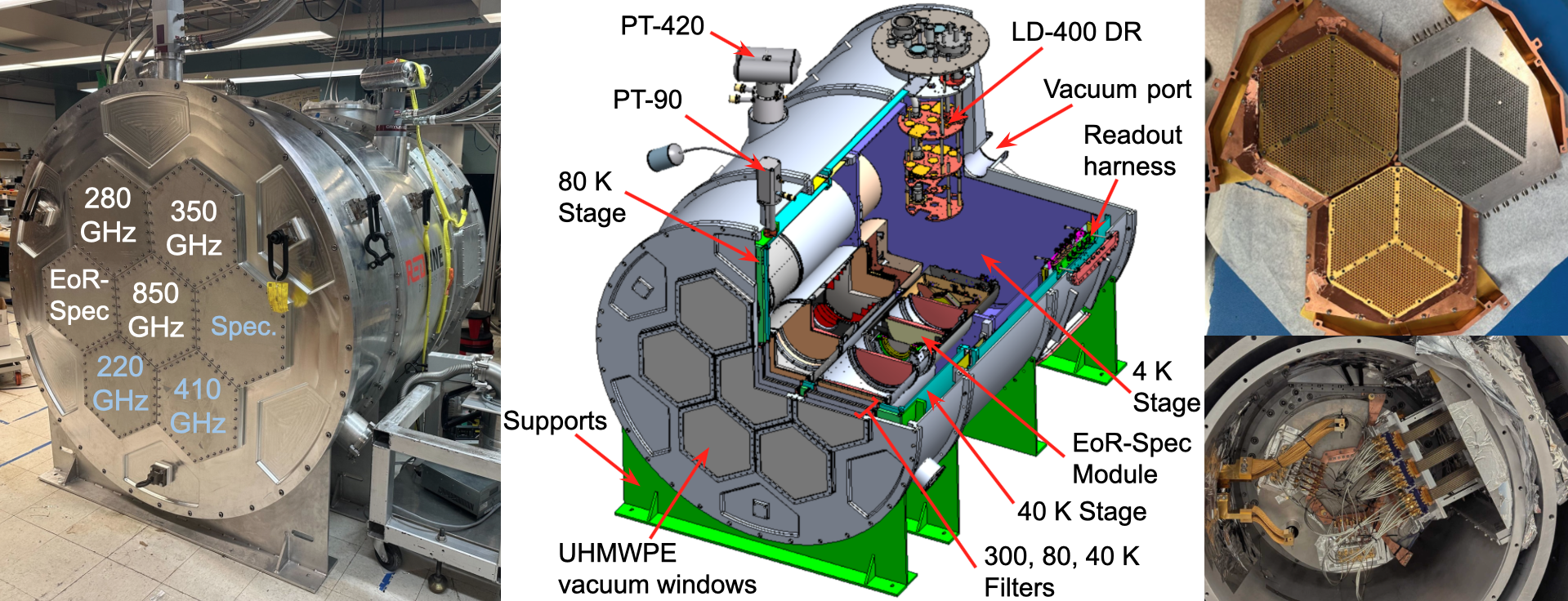}
\caption{{\it Left:} Prime-Cam in testing at Cornell with vacuum window blank-offs, and the planned positions of instrument modules labeled, with the first four labeled in white, and a preliminary additional three in light blue. {\it Center:} A cutaway CAD model of the Prime-Cam cryostat with instrument modules installed, showing optical elements and detector arrays. {\it Right, top:} The 280 GHz focal plane array, showing two gold-plated Si-platelet feedhorn assemblies for the Al arrays, and one Al-feedhorn assembly for the TiN arrays. {\it Right, bottom:} The 280 GHz module in testing in Mod-Cam, showing the cold readout from the back of the module through a readout harness, and the instrument module cold fingers and thermal straps connecting to the 1 K and 100 mK cold stages of the DR (not shown).
}
\label{fig:primecam}     
\end{figure}

The Prime-Cam cryostat was manufactured by Redline Chambers, and is currently undergoing testing at Cornell University (Fig. \ref{fig:primecam}). The cryostat includes aluminum 300 K, 80 K, 40 K and 4 K stages. The 4 K stage mechanically supports the instrument modules, while the 80 K and 40 K stages hold optical filters, and the 300 K stage holds optical filters and anti-reflection coated ultra high molecular-weight polyethylene vacuum windows. The stages are mechanically supported and thermally isolated from one another by a series of epoxied G10 tabs, and cooled by a Cryomech PT-90 and two PT-420 pulse tube cryocoolers. The 1 K and 100 mK instrument module stages are cooled by a Bluefors LD400 dilution refrigerator (DR). For more details on the Prime-Cam cryostat and optics design, see \cite{primecam2018} and \cite{primecam2024}.

\subsection{Instrument Modules} \label{sec:instmod}

Prime-Cam instrument modules contain optical elements and detectors, and are independently optimized and upgradable. Each contains 4 K, 1 K, and 100 mK stages. The modules are inserted into Prime-Cam from the rear and cantilevered off of the 4 K stage. The 4 K instrument module components include a metamaterial anti-reflection coated silicon lens and optical filters, as well as anti-reflection metamaterial tiles and magnetic shielding. The 1 K stage is supported off of the 4 K stage by an epoxied carbon fiber shell, and holds lenses, filters, and optical elements such as blackened baffles. The 100 mK stage is supported from the 1 K stage by a carbon fiber truss, and holds the detector array packages (Sec. \ref{sec:detectors}). 100 mK and 1 K instrument module cold fingers couple to the cold stages of the DR, and readout electronics extend from the back of the modules, to read out signals through Prime-Cam's three readout harnesses. For more details on the instrument module designs, see \cite{modcam2022}.

Funded instrument modules for Prime-Cam include the 280 \cite{modcam2022,modcam2025}, 350 \cite{primecam2024}, and 850 GHz \cite{850ghzchapman,850ghzhuber} broadband polarimetric instrument modules and the Epoch of Reionization spectrometer module (EoR-Spec) \cite{eorspec2024,eorspec2019}, which is equipped with a Fabry-Perot interferometer. Partially funded modules include a spectrometer-on-a-chip (spec-on-chip) module, and a 410 GHz module. The 280 GHz module will see first light in 2026, followed by the 350 GHz, 850 GHz, and EoR-Spec modules in 2027, and the spec-on-chip and 410 GHz modules thereafter. Modules may undergo testing in the Mod-Cam receiver (Sec. \ref{sec:modcam}) before deployment in Prime-Cam.

\subsection{Detectors and Readout} \label{sec:detectors}

Kinetic inductance detectors are an attractive superconducting detector technology for observations at millimeter and submillimeter wavelengths due to their natural frequency-division-multiplexed readout and relatively simple fabrication. Hundreds of KIDs can be read out on a single radio frequency (RF) transmission line, which requires sophisticated warm readout electronics and software control. To field the largest number of KIDs to date, Prime-Cam has necessitated significant developments in detector and readout technologies. More KIDs have been developed at the National Institute of Standards and Technology for Prime-Cam than have been deployed in previous experiments combined, and an extensive testing \cite{modcam2025,vaskuri2025,huber2024}, LED mapping \cite{middleton2024}, and characterization \cite{duell2024,Vaughan2025} efforts are underway for Prime-Cam's first four modules, which together will field over 64,000 total Al and TiN KIDs.

Detector timestream data is read out from Prime-Cam's instrument modules through custom readout harnesses, developed from the SO design \cite{mooreRH}. Each of Prime-Cam's readout harnesses is designed to accommodate 54 readout chains, with six chains per harness as spares \cite{Keller2025,Patel2025}. Prime-Cam uses custom-packaged FPGA-based Xilinx ZCU111 radio frequency system on chip (RFSoC) systems. Each RFSoC can simultaneously read out four RF channels with up to 1,000 detectors spanning a 512 MHz bandwidth per channel using the current firmware \cite{rfsoc}. Custom readout software interfaces with the boards via dedicated Ethernet lines, facilitating tasks such as frequency-multiplexed tone comb driving, comb calibration and optimization, and detector timestream establishment \cite{software}. 

\subsection{Mod-Cam Receiver} \label{sec:modcam}

Mod-Cam is a single-instrument-module testbed for Prime-Cam instrument modules and potential first light instrument for FYST, and is currently operating at Cornell University. Its off-axis DR design enables fast swapping of instrument modules through the rear of the cryostat, and includes 300 K, 40 K, and 4 K stages offset from one another using arrays of epoxied G10 tabs. Cooling power is supplied at 40 K and 4 K by a PT-420, and at 1 K and 100 mK by a Bluefors LD400 DR. For more details on Mod-Cam's design and cryogenic performance, see \cite{modcam2022,modcam2020,modcam2025}. Prime-Cam's 280 GHz module has been tested in dark, cold load, and optically open configurations inside of Mod-Cam (Fig. \ref{fig:primecam}) \cite{modcam2025}. Mod-Cam will serve as a testbed for future Prime-Cam modules.

\section{Summary and Status} \label{sec:conclusion}

FYST is currently under construction at the Cerro Chajnantor site. The FYST mirrors have been assembled and tested, and are en route to Chile for installation in 2026. Mod-Cam and the 280 GHz module have been tested, and are nearly ready for deployment. Prime-Cam is being developed and tested at Cornell, along with the 350 GHz module. EoR-Spec and the 850 GHz module are under construction, and the spec-on-chip and 410 GHz module are being planned. Detector, readout, and software development is ongoing.

Commissioning and early science on FYST will begin in 2026, and will include the 280 GHz, and potentially 350 GHz modules, fielding at minimum 10,000 KIDs. Science teams are readying analysis pipelines in anticipation of early science observations. Early activities include planetary calibration and studying and optimizing the wide field survey to prepare for full observations in the second year of observing, in addition to smaller, targeted field observations. Wide field observations will prioritize fields overlapping with SO observations, including CMB foreground and cluster observations in the Deep56 region \cite{actsurvey2014}, and/or the galactic plane. Early science goals include galaxy cluster studies, galaxy evolution in extragalactic fields, the characterization of CMB foregrounds and the atmosphere, galactic polarization, and time-domain science, including target-of-opportunity observations of transients. Observing with Prime-Cam on CCAT's FYST will usher in a new era of submillimeter observations for cosmology, astrophysics, and astronomy.

\end{document}